\documentclass[prd,aps,epsfig,amsmath,amssymb,nofootinbib,11pt]{revtex4}

\usepackage{epsfig}
\usepackage{graphicx}
\newcommand{\be}{\begin{equation}}
\newcommand{\ee}{\end{equation}}
\newcommand{\bea}{\begin{eqnarray}}
\newcommand{\beas}{\begin{eqnarray*}}
\newcommand{\eea}{\end{eqnarray}}
\newcommand{\eeas}{\end{eqnarray*}}
\newcommand{\ba}{\begin{array}}
\newcommand{\ea}{\end{array}}

\begin{document}
\title{Sneutrino Dark Matter in Light of PAMELA}   %%% Fill in title
\author{Rouzbeh Allahverdi}   %%% Fill in author names
\affiliation{Department of Physics \& Astronomy, University of New Mexico, Albuquerque, NM 87131, USA}    %%% Fill in author affiliations

\begin{abstract} %%% Abstract to run on from here.
In the $U(1)_{B-L}$ extension of the minimal supersymmetric standard model the right-handed sneutrino is a natural candidate for thermal dark matter. Sneutrino annihilation at the present time can be considerably enhanced due to the exchange of the lightest field in the Higgs sector that breaks $U(1)_{B-L}$. The annihilation mainly produces taus (or muons) by the virtue of $B-L$ charge assignments. A sneutrino mass of $1-2$~TeV provides a good fit to the PAMELA and is compatible with the latest results from the FERMI experiment. In addition, the sneutrino-nucleon elastic scattering cross section is
%$10^{-11}-10^{-9}$~pb, which might be probed by
within the reach of the upcoming and future direct detection experiments.
%In addition, if (at least) one of the neutrinos is dominantly a Dirac fermion, the sneutrino can provide a unified picture of dark matter and %inflation.
\end{abstract}

\maketitle



%%% MAIN BODY OF TEXT GOES HERE. CONSULT "INSTRUCTIONS FOR AUTHORS USING
%%% LATEX2E MARKUP", SECTIONS 2.3-2.6 FOR HELP WITH EQUATIONS, FIGURES,
%%% AND TABLES.

\section{Introduction}   %%% Top level section head (remove "%" symbol)

%Even though the existence of dark matter has been supported by various lines of evidence, the identity of dark matter itself is not yet known. One %proposed solution for this dark matter problem comes from particle physics beyond the standard model in the form of weakly interacting massive %particles (WIMPs)~\cite{Silk}.
There are currently major experimental efforts for direct and indirect detection of the dark matter particle.
The indirect detection investigates astrophysical effects of dark matter annihilation in the galaxy, including signatures in the cosmic rays.
%PAMELA is a satellite-borne experiment that measures cosmic ray fluxes.
The recently published results by PAMELA experiment show an excess of positron flux at energies above 10~GeV~\cite{PAMELA}, while no excess of anti-proton flux is observed~\cite{PAMELA-antiproton}.
%The publication shows results up to $\sim 100$~GeV and the experiment is expected to get data up to $\sim 190$~GeV for anti-protons and $\sim 270$~GeV %for positrons.
Another cosmic ray experiment called ATIC (a balloon experiment) has also recently published data where one observes an excess in the $e^{+} + e^{-}$ spectrum with a peak around 600~GeV~\cite{ATIC}.
%Another balloon experiment, the PPB-BETS~\cite{pep}, reports an excess in the $e^+ + e^-$ energy spectrum between 500 and 800 GeV.
%However, the excess is based on a few data-points that are not quite consistent with the ATIC data.
However, the latest results from the FERMI~\cite{FERMI} and H.E.S.S.~\cite{HESS} experiments do not confirm the peak at the high energies reported by ATIC.

While there could be astrophysical explanations for these anomalies (e.g. from nearby pulsars~\cite{pulsars}), it is reasonable to ask whether they can be attributed to the effect of dark matter annihilation in the galaxy.
Barring a large astrophysical boost factor $10^3-10^4$, which might be difficult to obtain based on recent analysis of halo substructure~\cite{halostructure}, a dark matter explanation requires an annihilation cross section much larger than the the canonical value $\sim 3 \times 10^{-26}$~cm$^3$/s~\cite{Vernon} and dominantly leptonic final states~\cite{Strumia}.
%Model-independent analysis shows that the annihilation cross section required to explain the positron excess exceeds the canonical value required by %relic density, i.e. $\sim 3 \times 10^{-26}$~cm$^3$/s, by at least an order of magnitude~\cite{Vernon}. In the usual neutralino dark matter scenario %in the minimal Supergravity (mSUGRA) model, the situation is further complicated because the dark matter annihilation cross section today is much %smaller than that at the freeze out time due to $P$-wave suppression.
%An astrophysical boost factor of $10^3-10^4$ is then needed to explain the observed positron excess~\cite{Lars}. However, this might be difficult to %obtain based on recent analysis of halo substructures (for example, see~\cite{halostructure}). Moreover, in order to explain both the positron and %anti-proton data, dark matter annihilation must be dominated by leptonic final state modes~\cite{Strumia,Salati}.
%(There could also be some effects from anisotropic propagation on the positron and anti-proton fluxes that still need to be %investigated~\citet{anisotropic}.)
This cannot be achieved for the neutralino dark matter in the minimal Supergravity (mSUGRA) model. There have been proposals for new dark matter models~\cite{Strumia,Nima} in which the dark matter candidate belongs to a hidden sector, and an acceptable thermal relic density is obtained via new gauge interactions.
The key ideas of these models are that the dark matter annihilation today is enhanced by a Sommerfeld effect~\cite{Sommerfeld} due to the existence of light bosons and that annihilation mainly produces lepton final states via symmetry of the hidden sector.
%This arrangement explains PAMELA data for a dark matter mass of a few hundred GeV, without needing a large astrophysical boost factor.
%and ATIC data for larger values of dark matter mass~\cite{Weiner}. Another type of explanation that has been proposed for the data is decaying dark %matter with a tuned lifetime~\cite{decaying}.

%We recently proposed an explicit model that can explain the measured anomalies in the cosmic rays~\cite{ADRS}. It is based on a simple extension of %the minimal supersymmetric standard model (MSSM) that includes a gauged $U(1)_{B-L}$ and where the dark matter is the lightest neutralino in the new %sector.
%Even though this model has a large dark matter annihilation cross section today due to Sommerfeld enhancement, the cross section for scattering of %dark matter off quarks is too low to be accessible to direct detection experiments. In fact, this is a generic situation for hidden sector dark matter %models that can explain PAMELA data along the line discussed above.

Here we consider an explicit model where dark matter belongs to the visible sector and can explain the positron excess. It is based on a simple extension of the minimal supersymmetric standard model (MSSM) that includes a gauged $U(1)_{B-L}$, with the right-handed (RH) sneutrino being the dark matter~\cite{ADRS2}.
%We argue that this is a minimal model of thermal dark matter that can explain the observed anomalies in cosmic rays and can also be probed by direct detection experiments.
%The main channel of sneutrino annihilation is to light Higgs fields, which carry a non-zero $B-L$ quantum number. These Higgs particles in turn decay %dominantly to leptons by virtue of the $B-L$ charges for fermions. The same Higgs field also results in a large Sommerfeld enhancement factor for the %annihilation cross section. For a sneutrino mass of 1-2 TeV, this model can explain PAMELA and ATIC data. In addition, due to the scalar nature of %dark matter, the sneutrino-proton elastic scattering cross section is in the $10^{-11}-10^{-9}$~pb range, which is an interesting range from a direct %detection perspective.
%Moreover, the sneutrino can be part of the field that drives primordial inflation, thus explaining the small temperature anisotropy in the cosmic %microwave background (CMB) via tiny neutrino masses~\cite{AKM,ADM}. We will also discuss various possibilities for radiative breaking of the $B-L$ %symmetry and some related issues.

\section{The model}

The $B-L$ extension of the MSSM~\citep{mohapatra} is well motivated since it automatically implies the existence of three RH neutrinos through which one can explain the neutrino masses and mixings. The minimal model contains a new gauge boson $Z^{\prime}$, two new Higgs fields $H^{\prime}_1$ and $H^{\prime}_2$, the RH neutrinos $N$, and their supersymmetric partners. The superpotential is (the boldface characters denote superfields)
\begin{equation} \label{sup}
W = W_{\rm MSSM} + W_{B-L} + y_D {\bf N}^c {\bf H_u} {\bf L} \, ,
\end{equation}
where ${\bf H_u}$ and ${\bf L}$ are the superfields containing the Higgs field that gives mass to up-type quarks and the left-handed (LH) leptons respectively. For simplicity, we have omitted the family indices. The $W_{B-L}$ term contains ${\bf H^{\prime}_1},~{\bf H^{\prime}_2}$, ${\bf N}^c$ and its detailed form depends on the charge assignments of the new Higgs fields. The last term on the RH side of Eq.~(\ref{sup}) is the neutrino Yukawa coupling term.

%The scalar potential consists of $F$-terms from the superpotential, and $D$-terms from the gauge symmetries.
%The $D$-term contribution from $U(1)_{B-L}$ is given by
%
%\begin{equation} \label{dterm1}
%V_D \supset \frac{1}{2} D^2_{B-L} ,
%\end{equation}
%
%where
%
%\begin{equation} \label{dterm2}
%D_{B-L} = \frac{1}{2} \gBL \left[Q_{1} ({\vert H^{\prime}_1 \vert}^2 - {\vert H^{\prime}_2 \vert}^2) + \frac{1}{2} {\vert \tilde N \vert}^2 + ... %\right] .
%\end{equation}
%
%Here $\gBL$ is the gauge coupling of $U(1)_{B-L}$, and $+Q_1$, $-Q_1$, $1/2$ are the $B-L$ charges of $H^{\prime}_1,~H^{\prime}_2,~{\tilde N}$ %respectively (${\tilde N}$ is the sneutrino field).
The $U(1)_{B-L}$ is broken by the vacuum expectation value (VEV) of $H^{\prime}_1$ and $H^{\prime}_2$, which we denote by $v^{\prime}_1$ and $v^{\prime}_2$ respectively. This results in a mass $m_{Z^\prime} = g_{B-L} Q_1 \sqrt{v^{\prime 2}_1 + v^{\prime 2}_2}$ for the $Z^{\prime}$ gauge boson. Here $g_{B-L}$ is the gauge coupling of $U(1)_{B-L}$, and $+Q_1$, $-Q_1$ are the $B-L$ charges of $H^{\prime}_1,~H^{\prime}_2$ respectively. We have three physical Higgs fields $\phi,~\Phi$ (scalars) and ${\cal A}$ (a pseudo scalar). The scalar Higgses are related to the real parts of $H^{\prime}_1,~H^{\prime}_2$ through the mixing angle $\alpha^{\prime}$:
\begin{eqnarray} \label{vev}
H^{\prime}_1 & = & \frac{v^{\prime}_1 + \cos \alpha^{\prime} \Phi - \sin \alpha^{\prime} \phi}{\sqrt{2}} + \frac{H^{\prime}_{1,I}}{\sqrt{2}}  \, \nonumber \\
H^{\prime}_2 & = & \frac{v^{\prime}_2 + \sin \alpha^{\prime} \Phi + \cos \alpha^{\prime} \phi}{\sqrt{2}} + \frac{H^{\prime}_{2,I}}{\sqrt{2}}  \, ,
\end{eqnarray}
where $H^{\prime}_{1,I}, H^{\prime}_{2,I}$ represent the imaginary parts.
%Eqs.~(\ref{dterm1},\ref{dterm2},\ref{vev}) lead to the following terms in the scalar potential
%
%\begin{eqnarray} \label{Ncoupling}
%V &\supset& -\frac{1}{2} \gBL m_{Z^\prime} \sin (\alpha^{\prime} + \beta^{\prime}) \, \phi \, {\vert \tilde N \vert}^2 \, \nonumber \\
%&& - \frac{1}{2} \gBL^{2} Q_1 \cos (2 \alpha^{\prime}) \, \phi^2 \, {\vert \tilde N \vert}^2 \, \nonumber \\
%&& + \frac{1}{2} \gBL m_{Z^\prime} \cos (\alpha^{\prime} + \beta^{\prime}) \, \Phi \, {\vert \tilde N \vert}^2 \, \nonumber \\
%&& + \ ... \ ,
%\end{eqnarray}
%
%where $\tan \beta^{\prime} \equiv v^{\prime}_2/v^{\prime}_1$.
The masses of the Higgs fields follow $m^2_{\phi} < \cos^2 (2 \beta^{\prime}) \, m^2_{Z^\prime}$ and $m_{\Phi},~m_{\cal A} \sim m_{Z^\prime}$ ($\tan \beta^{\prime} \equiv v^{\prime}_2/v^{\prime}_1$).

A natural dark matter candidate in this model is the sneutrino ${\tilde N}$~\footnote{Another candidate is the lightest neutralino in the new sector~\cite{ADRS}.
%, which is a linear combination of the $U(1)_{B-L}$ gaugino ${\widetilde Z}^{\prime}$ and the two Higgsinos ${\widetilde H}^{\prime}_1$, ${\widetilde H}^{\prime}_2$~\cite{ADRS}.
%However, we can make this neutralino heavier than the sneutrino by choosing the gaugino and Higgsino mass parameters appropriately. The sneutrino mass %is fixed by the scalar soft mass terms.
}.
% since its mass receives the smallest contribution from the gaugino loops.
The main processes for annihilation of dark matter quanta are then governed by the $D$-term contribution to the scalar potential~\cite{ADRS2}, with the dominant mode being ${\tilde N}^* {\tilde N} \rightarrow \phi \phi$.
%via the $s$-channel exchange of the $\phi,~\Phi$, the $t$,~$u$-channel exchange of the ${\tilde N}$, and the contact term $\vert {\tilde N} \vert^2 %\phi^2$~\footnote{
%The $s$-channel $Z^\prime$ exchange is subdominant because of the large $Z^\prime$ mass (as required by the experimental bound on $m_{Z^\prime}$).
%There are other annihilation modes with Higgs or RH neutrino final states but
%${\tilde N}^* {\tilde N} \rightarrow \phi \Phi , ~ \phi {\cal A}, ~ \Phi \Phi,~{\cal A} {\cal A}$ annihilation processes, but
%they are kinematically forbidden o suppressed for the parameter space that we consider.}
%the annihilation into $\nu \bar{\nu}$ final states is at least an order of magnitude below the $\phi \phi$ final states. Other fermion final states, %through $s$-channel $Z^\prime$ exchange, have even smaller branching ratios. Moreover, note that the annihilations to fermion-antifermion final states %are $P$-wave suppressed.}.
The $\phi$ subsequently decays into fermion-antifermion pairs via a one-loop diagram containing two $Z^{\prime}$ bosons. The decay rate is given by:
\begin{equation} \label{phidec}
\Gamma(\phi \to f {\bar f}) = \frac{C_f}{2^7 \pi^5} \frac{g_{B-L}^6 Q^4_{f} Q^2_{\phi} m^5_{\phi} m^2_f}{m^6_{Z^{\prime}}} \left(1 - \frac{4 m^2_f}{m^2_{\phi}} \right)^{3/2},
\end{equation}
where $Q_f$ and $Q_\phi$ are the $B-L$ charges of the final state fermion and the $\phi$ respectively, $m_f$ is the fermion mass, and $C_f$ denotes color factor.
Since the $B-L$ charge of leptons is three times larger than that of quarks, the leptonic branching ratio is naturally larger than that for quarks.
%We note that $m_{\phi}$ can be controlled by the VEVs of the new Higgs fields and for comparable VEVs, i.e.
For $\tan \beta^{\prime} \approx 1$, we can have $m_{\phi} \ll m_{Z^\prime}$. If $m_{\phi} > 2 \, m_b$, the dominant decay mode is $\phi \rightarrow \tau^{-} \tau^{+}$ final state, while the branching ratio for the $\phi \rightarrow b {\bar b}$ mode is $\approx 7$ times smaller.

The annihilation cross section at the present time has Sommerfeld enhancement as a result of the attractive force between sneutrinos due to the $\phi$ exchange
%. The Higgs coupling to dark matter is given by the first term on the RH side of Eq.~(\ref{Ncoupling}) and
that leads to an attractive potential $V(r) = -\alpha (e^{-m_{\phi}r}/r)$ in the non-relativistic limit~\citep{Sommerfeld}, where
\begin{equation} \label{fsc}
\alpha = \frac{g_{B-L} m_{Z^\prime} \sin (\alpha^{\prime} + \beta^{\prime})}{4 m_{\tilde N}} \ ,
\end{equation}
and $m_{\tilde N}$ is the sneutrino mass.

\section{Sneutrino dark matter and PAMELA}

%We are now going to show that the sneutrino dark matter can explain the PAMELA data.
%We first identify the allowed regions of the model parameter space that result in an acceptable dark matter relic density and then find the Sommerfeld %enhancement factor for these regions.
As an explicit example,
%which we call the minimal model,
we choose the $B-L$ charge for $H_1^\prime$ (i.e. $Q_1$) to be $3/2$. The $B-L$ charges of quarks and leptons are chosen to be $1/6$ and $-1/2$ respectively.
%The $B-L$ charges of the fields involved are shown in Table~\ref{BLcharges-tbl}.
%
%\begin{table}[!ht]
%\caption{The $B-L$ charges of the fields for the minimal model. Here $Q$ and $L$ represent quarks and leptons respectively, while $H^{\prime}_1$ and $H^{\prime}_2$ are the two new Higgs fields. The MSSM Higgs fields have zero $B-L$ charges.}
%\smallskip
%\begin{center}
%{\small
%\begin{tabular}{ccccccc}
%\tableline
%\noalign{\smallskip}
%{\rm Fields} & $Q$ & $Q^c$ & $L$ & $L^c$ & $H^{\prime}_1$ & $H^{\prime}_2$ \\ \hline
%\noalign{\smallskip}
%\tableline
%\noalign{\smallskip}
%$Q_{B-L}$ & 1/6 & -1/6 & -1/2 & 1/2 & 3/2 & -3/2 \\ \hline
%2 & $-599$  & $4.1 \times 10^{15}$ & $2.1 \times 10^{15}$ & 0.9\\
%3 & $-792$  & $2.1 \times 10^{15}$ & $1.7 \times 10^{15}$ & 0.7\\
%4 & $-1029$ & $6.0 \times 10^{15}$ & $2.3 \times 10^{15}$ & 0.5\\
%\noalign{\smallskip}
%\tableline
%\end{tabular}
%}
%\end{center}
%\label{BLcharges-tbl}
%\end{table}

%\begin{table}[tbp]
%\center
%\begin{tabular}{|c||c|c|c|c|c|c|c|}\hline
%{\rm Fields} & $Q$ & $Q^c$ & $L$ & $L^c$ &  $H^{\prime}_1$ & $H^{\prime}_2$ \\ \hline
%$\QBL$ & 1/6 & -1/6 & -1/2 & 1/2 &  3/2 & -3/2 \\ \hline
%\end{tabular}
%\caption{The $B-L$ charges of the fields for the minimal model. Here $Q$ and $L$ represent quarks and leptons respectively, while $H^{\prime}_1$ and %$H^{\prime}_2$ are the two new Higgs fields. The MSSM Higgs fields have zero $B-L$ charges.}
%\label{BLcharges-tbl}
%\end{table}
%
We use reasonable values for the model parameters, i.e., $\tan \beta^\prime \approx 1$, $m_{Z^\prime} > 1.5$~TeV, $\mu^\prime = 0.5 - 1.5$~TeV ($\mu^{\prime}$ being the Higgs mixing parameter in the $B-L$ sector), soft masses for the Higgs fields $m_{H_{1,2}^\prime} = 200-600$~GeV, and soft gaugino mass $M_{\widetilde{Z}^\prime} \geq 1$~TeV.
%We use $g_{B-L} \sim 0.40$, which is in concordance with unification of the gauge couplings (we need to use a normalization factor $\sqrt{3/2}$ for %unification).
%We show the unification of all the gauge couplings using the two loop renormalization group equations (RGEs) in Figure~\ref{unification}.
%We find that for $m_{Z^{\prime}} \simeq 2.5$ TeV the couplings unify at $\sim 10^{16}$ GeV. This figure is drawn for the $B-L$ assignments shown in %Table~\ref{BLcharges-tbl}.
The $Z^{\prime}$ mass used in the calculation obeys the LEP and Tevatron bounds~\citep{Z'} for our charge assignments. The sneutrino mass is chosen to be between 800 GeV and 2 TeV in order to explain the PAMELA data.

We
%select points that satisfy the dark matter relic density and then
use {\tt DarkSUSY}-5.0.2~\cite{darksusy} to calculate the positron flux from dark matter annihilation. Each pair annihilation in our model produces 2 $\phi$'s that yield four fermions upon their decay. For this reason, we generally need a heavier sneutrino compared to models in which the pair annihilation directly produces fermions. We normalize the positron fraction by a factor $k_b = 1.11$ according to~\cite{BE}.
%There are theoretical uncertainties in the positron cosmic ray flux calculation due to the assumptions about the dark matter halo profile and the %cosmic ray propagation model.
Here we assume NFW profile~\cite{NFW} for the dark matter halo and MED parameters for the propagation as defined in~\cite{Delahaye}.

\begin{figure}[ht]
\vspace*{1.0cm}
\begin{center}
\vskip -0.9in
\includegraphics[width=.48\textwidth]{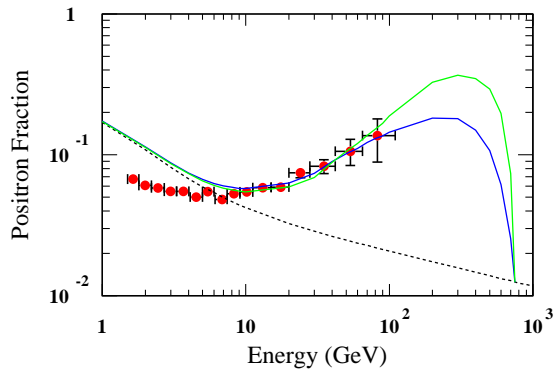}
\end{center}
\vskip -0.65in
\caption{We show a fit to the PAMELA data when the $\phi$ decays mostly to taus (dark blue) or muons (light green) for a sneutrino mass of 1.5~TeV and an enhancement factor of $10^3$. The dashed line is the expected background cosmic rays.}
\label{pamelatau}
\end{figure}

In Figure~\ref{pamelatau}, we show our fit to the PAMELA data for $m_{\tilde N} = 1.5$~TeV for $\tau^+ \tau^-$ and $\mu^+ \mu^-$ final state cases. We found that with an enhancement factor of $10^3$ the chi-square values (including only points with energy greater than 10~GeV) for a sneutrino mass of 1.5~TeV are small, i.e. 2.9 and 5.5 for $\tau^+ \tau^-$ and $\mu^+ \mu^-$ respectively.
%When $m_{\phi}$ is (chosen to be) below $2 \, m_b$ but above $2 \, m_{\tau}$, we do not have any anti-proton excess.
We can raise $m_{\phi}$ up to $\sim 15$~GeV and still have acceptable anti-proton flux.
Also, we find that for $\tau^+ \tau^-$ final state the $e^+ + e^-$ spectrum at higher energies is compatible with the recent results from FERMI satellite.
%have a reasonable fit to the ATIC data, although simultaneous fit for both ATIC and PAMELA are not satisfactory~\cite{ADRS}.
%We note that this model also has a great potential to be observed with the Fermi Satellite experiment. Due to electromagnetically charged final states %of $\phi$ decays, the Sommerfeld enhancement would also lead to a higher rate of photons in the gamma ray background~\cite{CRgamma}. There could also %be contribution to the neutrino flux~\cite{CRneutrino}.

%\begin{figure}[!h]
%\plotone{pamela-taumu-S1k.eps}
%\includegraphics{We show a fit to the PAMELA data when the $\phi$ decays mostly to taus (dark blue) or muons (light green) for a sneutrino mass of %1.5~TeV and an enhancement factor of $10^3$. The dashed line is the expected background cosmic rays.}
%\label{pamelatau}
%\end{figure}

\section{Direct detection}
\begin{figure}[ht]
\vspace*{1.0cm}
\begin{center}
\vskip -0.4in
\includegraphics[width=.48\textwidth]{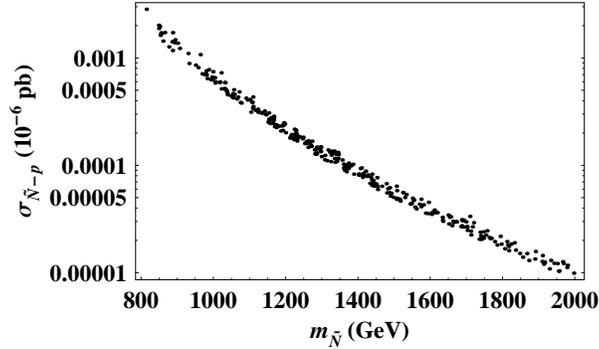}
\end{center}
\vskip -0.25in
\caption{We show the direct detection cross section as a function of sneutrino mass.}
\label{direct}
\end{figure}

%The current upper bound on the spin-independent dark matter particle-proton scattering cross section is about $4.6 \times 10^{-8}$~pb for a dark %matter mass around 60~GeV, and increasing to $\sim 2 \times 10^{-7}$~pb for a mass around 1.2~TeV~\citep{cdms}.
In our model the elastic scattering of the sneutrino occurs via the $Z^{\prime}$ exchange with the nucleus in the $t$-channel. This leads to only a spin-independent contribution since the $B-L$ charges of the left and right quarks are the same. In Figure~\ref{direct}, we show the $\tilde N$-$p$ scattering cross section for the model points that satisfy the relic density constraint $0.096 < \Omega_{DM} h^2 < 0.124$~\cite{WMAP5}. We see that the cross section can be in the $10^{-11}-10^{-9}$ pb range, which is close to the reach of the upcoming dark matter direct detection experiments~\citep{Direct}. This also gives rise to neutrino signals from dark matter annihilation that are detectable at the IceCube neutrino telescope~\cite{abdr}.

%It is also seen that the cross section decreases as the sneutrino mass increases. This is because for larger values of $m_{\tilde N}$ we also need a %larger annihilation cross section to satisfy the relic density constraint. As discussed earlier the annihilation cross section depends on the %sneutrino couplings to $\phi$ and $\Phi$, which are $\propto m_{Z^\prime}$.

%\begin{figure}[!h]
%\includegraphics{directsneutrino.eps}
%\caption{We show the direct detection cross section as a function of sneutrino mass.}
%\label{direct}
%\end{figure}

\section{Acknowledgements} %%% Text of acknowledgements runs on after this command.

%I would like to thank the organizers of SUSY 2009 Conference for their invitation and kind hospitality.
I would like to thank Bhaskar Dutta, Katherine Richardson-Mcdaniel, Yudi Santoso and Sascha Bornhauser for collaboration and numerous discussions on various aspects of this subject.

%%% THE BIBLIOGRAPHY
%%%
%%% CONSULT SECTION 3 OF "INSTRUCTIONS FOR AUTHORS" FOR HOW TO USE NATBIB.
%%% AUTHORS ARE ENCOURAGED TO USE EITHER THE "THEBIBLIOGRAPY" ENVIRONMENT
%%% BY UNCOMMENTING (DELETING THE "%" SYMBOL) THE COMMANDS BELOW, OR BY
%%% USING THE BIBTEX ENVIRONMENT. TO FIND OUT WHICH IS APPLICABLE TO YOUR
%%% CONTRIBUTION, CONSULT THE VOLUME EDITORS FOR YOUR PROCEEDINGS.
%%%

\end{document}